\documentclass{singlecol-new}

\usepackage{array}
\usepackage{algorithm}
\usepackage{algorithmic}
\usepackage{url}
\usepackage{ifpdf}
\usepackage{graphicx}

\usepackage{natbib,stfloats}
\usepackage{mathrsfs}

\theoremstyle{TH}{

}

\theoremstyle{THrm}{

}

\theoremstyle{THhit}{

}

\makeatletter
\def\theequation{\arabic{equation}}
\makeatother
\begin{document}

\title{Vulnerability Detection in C/C++ Code with Deep Learning}

\authorA{Zhen Huang}
\affA{School of Computing, \\DePaul University, \\Chicago, IL, USA \\
\qquad E-mail: zhen.huang@depaul.edu}

\authorB{Amy Aumpansub}
\affB{School of Computing, \\DePaul University, \\Chicago, IL, USA \\
\qquad E-mail: amy.aumpansub@gmail.com}

\begin{abstract}
Deep learning has been shown to be a promising tool in detecting software vulnerabilities. In this work, we train neural networks with program slices extracted from the source code of C/C++ programs to detect software vulnerabilities. The program slices capture the syntax and semantic characteristics of vulnerability-related program constructs, including API function call, array usage, pointer usage, and arithmetic expression. To achieve a strong prediction model for both vulnerable code and non-vulnerable code, we compare different types of training data, different optimizers, and different types of neural networks. Our result shows that combining different types of characteristics of source code and using a balanced number of vulnerable program slices and non-vulnerable program slices produce a balanced accuracy in predicting both vulnerable code and non-vulnerable code. Among different neural networks, BGRU with the ADAM optimizer performs the best in detecting software vulnerabilities with an accuracy of 92.49\%.
\end{abstract}

\KEYWORD{software vulnerabilities; vulnerability detection; deep learning; neural networks; program analysis.}

\maketitle
\pagestyle{empty}
\thispagestyle{empty}

\newcommand{\nmappings}{11314}
\newcommand{\napps}{1260}





\newcommand{\myfig}[5]
{
\begin{figure}[!ht]
\begin{center}
\ifpdf
\includegraphics[width=#4\linewidth]{#1}
\else
\includegraphics[width=#4\linewidth]{#1}
\fi
\end{center}
\vspace{#5}
\caption{#2}\label{#3}
\end{figure}
}

\newcommand{\myfigwide}[5]
{
\begin{figure*}[t]
\begin{center}
\ifpdf
\includegraphics[width=#4\linewidth]{#1}
\else
\includegraphics[width=#4\linewidth]{#1}
\fi
\end{center}
\vspace{#5}
\caption{#2}\label{#3}
\end{figure*}
}

\newcommand{\mysubfigtwo}[8]
{
\begin{figure}
        \centering
        \begin{subfigure}[b]{0.25\textwidth}
                \centering
                \includegraphics[width=\textwidth]{#1}
                \caption{#2}
                \label{#3}
        \end{subfigure}%
        ~ 
        \begin{subfigure}[b]{0.25\textwidth}
                \centering
                \includegraphics[width=\textwidth]{#4}
                \caption{#5}
                \label{#6}
        \end{subfigure}
        ~ 
        \caption{#7}\label{#8}
        \vspace{-15pt}        
\end{figure}
}

\newcommand{\myfigthreevert}[9]
{
\def\tempa{#1}
\def\tempb{#2}
\def\tempc{#3}
\def\tempd{#4}
\def\tempe{#5}
\def\tempf{#6}
\def\tempg{#7}
\def\temph{#8}
\def\tempi{#9}
\myfigthreevertcont
}

\newcommand{\myfigthreevertcont}[2]
{
\begin{figure}
        \centering
         \begin{tabular}{c}
        \begin{subfigure}[b]{0.6\textwidth}
                \centering
                \includegraphics[width=\textwidth]{\tempa}
                \caption{\tempb}
                \label{\tempc}
        \end{subfigure}%
	\\
	\\
	~
        \begin{subfigure}[b]{0.6\textwidth}
                \centering
                \includegraphics[width=\textwidth]{\tempd}
                \caption{\tempe}
                \label{\tempf}
        \end{subfigure}%
	\\
	\\
	~
        \begin{subfigure}[b]{0.6\textwidth}
                \centering
                \includegraphics[width=\textwidth]{\tempg}
                \caption{\temph}
                \label{\tempi}
        \end{subfigure}%
	\\
	\\
	\end{tabular}
	~
	\caption{#1}\label{#2}
        \vspace{-5pt}        
\end{figure}
}
\newcommand{\myfigthreevertbox}[9]
{
\def\tempa{#1}
\def\tempb{#2}
\def\tempc{#3}
\def\tempd{#4}
\def\tempe{#5}
\def\tempf{#6}
\def\tempg{#7}
\def\temph{#8}
\def\tempi{#9}
\myfigthreevertboxcont
}

\newcommand{\myfigthreevertboxcont}[2]
{
\begin{figure}
        \centering
         \begin{tabular}{c}
        \begin{subfigure}[b]{0.6\textwidth}
                \centering
                \setlength{\fboxsep}{0pt}%
                \setlength{\fboxrule}{1pt}%
                \fbox{\includegraphics[width=\textwidth]{\tempa}}%
                \caption{\tempb}
                \label{\tempc}
        \end{subfigure}%
	\\
	\\
	~
        \begin{subfigure}[b]{0.6\textwidth}
                \centering
                \setlength{\fboxsep}{0pt}%
                \setlength{\fboxrule}{1pt}%
                \fbox{\includegraphics[width=\textwidth]{\tempd}}%
                \caption{\tempe}
                \label{\tempf}
        \end{subfigure}%
	\\
	\\
	~
        \begin{subfigure}[b]{0.6\textwidth}
                \centering
                \setlength{\fboxsep}{0pt}%
                \setlength{\fboxrule}{1pt}%
                \fbox{\includegraphics[width=\textwidth]{\tempg}}%
                \caption{\temph}
                \label{\tempi}
        \end{subfigure}%
	\\
	\\
	\end{tabular}
	~
	\caption{#1}\label{#2}
        \vspace{-5pt}        
\end{figure}
}

\newcommand{\myfigtwovert}[8]
{
\begin{figure}
        \centering
         \begin{tabular}{c}
        \begin{subfigure}[b]{0.6\textwidth}
                \centering
                \includegraphics[width=\textwidth]{#1}
                \caption{#2}
                \label{#3}
        \end{subfigure}%
        \\
	\\
        ~ 
        \begin{subfigure}[b]{0.6\textwidth}
                \centering
                \includegraphics[width=\textwidth]{#4}
                \caption{#5}
                \label{#6}
        \end{subfigure}
        \end{tabular}
        ~ 
        \caption{#7}\label{#8}
        \vspace{-10pt}        
\end{figure}
}

\newcommand{\mysubfigtwovert}[8]
{
\begin{figure}
         \begin{tabular}{c}
        \begin{subfigure}[b]{0.3\textwidth}
                \centering
                \includegraphics[width=\textwidth]{#1}
                \caption{#2}
                \label{#3}
        \end{subfigure}%
        \\
        ~ 
        \begin{subfigure}[b]{0.3\textwidth}
                \centering
                \includegraphics[width=\textwidth]{#4}
                \caption{#5}
                \label{#6}
        \end{subfigure}
        \end{tabular}
        ~ 
        \caption{#7}\label{#8}
        \vspace{-10pt}        
\end{figure}
}

\newcommand{\mysubfigthree}[9]
{
\def\tempa{#1}
\def\tempb{#2}
\def\tempc{#3}
\def\tempd{#4}
\def\tempe{#5}
\def\tempf{#6}
\def\tempg{#7}
\def\temph{#8}
\def\tempi{#9}
\mysubfigthreecont
}

\newcommand{\mysubfigthreecont}[2]
{
\begin{figure*}
        \centering
        \begin{subfigure}[b]{0.3\textwidth}
                \centering
                \includegraphics[width=\textwidth]{\tempa}
                \caption{\tempb}
                \label{\tempc}
        \end{subfigure}%
        ~ 
        \begin{subfigure}[b]{0.3\textwidth}
                \centering
                \includegraphics[width=\textwidth]{\tempd}
                \caption{\tempe}
                \label{\tempf}
        \end{subfigure}
        \begin{subfigure}[b]{0.3\textwidth}
                \centering
                \includegraphics[width=\textwidth]{\tempg}
                \caption{\temph}
                \label{\tempi}
        \end{subfigure}
        ~ 
        \caption{#1}\label{#2}
        \vspace{-5pt}
\end{figure*}
}

\newcommand{\mysubfigthreebox}[9]
{
\def\tempa{#1}
\def\tempb{#2}
\def\tempc{#3}
\def\tempd{#4}
\def\tempe{#5}
\def\tempf{#6}
\def\tempg{#7}
\def\temph{#8}
\def\tempi{#9}
\mysubfigthreeboxcont
}

\newcommand{\mysubfigthreeboxcont}[2]
{
\begin{figure*}
        \centering
        \begin{tabular}{ccc}
        \begin{subfigure}[b]{0.32\textwidth}
                \centering
                {%
                \setlength{\fboxsep}{0pt}%
                \setlength{\fboxrule}{1pt}%
                \fbox{\includegraphics[width=\textwidth]{\tempa}}%
                }%
                \caption{\tempb}
                \label{\tempc}
        \end{subfigure}%
        &
        \begin{subfigure}[b]{0.32\textwidth}
                \centering
                {%
                \setlength{\fboxsep}{0pt}%
                \setlength{\fboxrule}{1pt}%
                \fbox{\includegraphics[width=\textwidth]{\tempd}}%
                }%
                \caption{\tempe}
                \label{\tempf}
        \end{subfigure}
        &
        \begin{subfigure}[b]{0.32\textwidth}
                \centering
                {%
                \setlength{\fboxsep}{0pt}%
                \setlength{\fboxrule}{1pt}%
                \fbox{\includegraphics[width=\textwidth]{\tempg}}%
                }%
                \caption{\temph}
                \label{\tempi}
        \end{subfigure}
        \end{tabular}
        \caption{#1}\label{#2}
        \vspace{-10pt}
\end{figure*}
}

\newcommand{\mysubfigfourbox}[9]
{
\def\tempa{#1}
\def\tempb{#2}
\def\tempc{#3}
\def\tempd{#4}
\def\tempe{#5}
\def\tempf{#6}
\def\tempg{#7}
\def\temph{#8}
\def\tempi{#9}
\mysubfigfourboxcont
}

\newcommand{\mysubfigfourboxcont}[5]
{
\begin{figure*}
        \centering
        \begin{tabular}{cc}
        \begin{subfigure}[b]{0.5\textwidth}
                \centering
                {%
                \setlength{\fboxsep}{0pt}%
                \setlength{\fboxrule}{1pt}%
                \fbox{\includegraphics[width=\textwidth]{\tempa}}%
                }%
                \caption{\tempb}
                \label{\tempc}
        \end{subfigure}%
        &
        \begin{subfigure}[b]{0.5\textwidth}
                \centering
                {%
                \setlength{\fboxsep}{0pt}%
                \setlength{\fboxrule}{1pt}%
                \fbox{\includegraphics[width=\textwidth]{\tempd}}%
                }%
                \caption{\tempe}
                \label{\tempf}
        \end{subfigure}
        \\
        \begin{subfigure}[b]{0.5\textwidth}
                \centering
                {%
                \setlength{\fboxsep}{0pt}%
                \setlength{\fboxrule}{1pt}%
                \fbox{\includegraphics[width=\textwidth]{\tempg}}%
                }%
                \caption{\temph}
                \label{\tempi}
        \end{subfigure}
        &
        \begin{subfigure}[b]{0.5\textwidth}
                \centering
                {%
                \setlength{\fboxsep}{0pt}%
                \setlength{\fboxrule}{1pt}%
                \fbox{\includegraphics[width=\textwidth]{#1}}%
                }%
                \caption{#2}
                \label{#3}
        \end{subfigure}
        \end{tabular}
        \caption{#4}\label{#5}
        \vspace{-10pt}
\end{figure*}
}

\newcommand{\mysubfigsixbox}[9]
{
\def\tempa{#1}
\def\tempb{#2}
\def\tempc{#3}
\def\tempd{#4}
\def\tempe{#5}
\def\tempf{#6}
\def\tempg{#7}
\def\temph{#8}
\def\tempi{#9}
\mysubfigsixboxcont
}

\newcommand{\mysubfigsixboxcont}[9]
{
\def\tempj{#1}
\def\tempk{#2}
\def\templ{#3}
\def\tempm{#4}
\def\tempn{#5}
\def\tempo{#6}
\def\tempp{#7}
\def\tempq{#8}
\def\tempr{#9}
\mysubfigsixboxcontcont
}

\newcommand{\mysubfigsixboxcontcont}[2]
{
\begin{figure*}
        \centering
        \begin{tabular}{ccc}
        \begin{subfigure}[b]{0.32\textwidth}
                \centering
                {%
                \setlength{\fboxsep}{0pt}%
                \setlength{\fboxrule}{1pt}%
                \fbox{\includegraphics[width=\textwidth]{\tempa}}%
                }%
                \caption{\tempb}
                \label{\tempc}
        \end{subfigure}%
        &
        \begin{subfigure}[b]{0.32\textwidth}
                \centering
                {%
                \setlength{\fboxsep}{0pt}%
                \setlength{\fboxrule}{1pt}%
                \fbox{\includegraphics[width=\textwidth]{\tempd}}%
                }%
                \caption{\tempe}
                \label{\tempf}
        \end{subfigure}
        &
        \begin{subfigure}[b]{0.32\textwidth}
                \centering
                {%
                \setlength{\fboxsep}{0pt}%
                \setlength{\fboxrule}{1pt}%
                \fbox{\includegraphics[width=\textwidth]{\tempg}}%
                }%
                \caption{\temph}
                \label{\tempi}
        \end{subfigure}
        \\
        \begin{subfigure}[b]{0.32\textwidth}
                \centering
                {%
                \setlength{\fboxsep}{0pt}%
                \setlength{\fboxrule}{1pt}%
                \fbox{\includegraphics[width=\textwidth]{\tempj}}%
                }%
                \caption{\tempk}
                \label{\templ}
        \end{subfigure}
        &
        \begin{subfigure}[b]{0.32\textwidth}
                \centering
                {%
                \setlength{\fboxsep}{0pt}%
                \setlength{\fboxrule}{1pt}%
                \fbox{\includegraphics[width=\textwidth]{\tempm}}%
                }%
                \caption{\tempn}
                \label{\tempo}
        \end{subfigure}
        &
        \begin{subfigure}[b]{0.32\textwidth}
                \centering
                {%
                \setlength{\fboxsep}{0pt}%
                \setlength{\fboxrule}{1pt}%
                \fbox{\includegraphics[width=\textwidth]{\tempp}}%
                }%
                \caption{\tempq}
                \label{\tempr}
        \end{subfigure}
        \end{tabular}
        \vspace{-8pt}
        \caption{#1}\label{#2}
\end{figure*}
}

\newcommand{\BA}{{\em begin\_atomic}}
\newcommand{\EA}{{\em end\_atomic}}

\newcounter{claimcounter}[section]
\newcolumntype{R}[1]{>{\raggedleft\arraybackslash}m{#1}}
\newcommand{\scbf}[1]{\vspace {0.05in}\noindent{\textbf{#1.}}}

\newcommand{\algorithmicbreak}{\textbf{break}}
\newcommand{\Break}{\State \algorithmicbreak}
\newcommand{\algorithmiccontinue}{\textbf{continue}}
\newcommand{\Continue}{\State \algorithmiccontinue}



\section{Introduction}
Software vulnerabilities pose a significant threat to the security of networks and information. Hackers and malware often take advantage of these vulnerabilities to compromise computer systems, because vulnerabilities enable them to  dramatically increase the magnitude and speed of cyber attacks. To incentivize individuals to find such vulnerabilities, renowned software vendors are known to offer rewards as high as \$1 million~\citep{intel-bugbounty, microsoft-bugbounty, apple-bugbounty, facebook-bugbounty}.


As manually finding vulnerabilities typically incur considerable effort and time, numerous studies have been dedicated to automatically identify vulnerabilities~\citep{Vuddy, VulPecker, VDiscover, VulnerableComponents, Chucky, VulnerabilityExtrapolation}. Primarily, these approaches rely on code similarity detection or pattern matching techniques. However, code similarity detection may not effectively identify vulnerabilities that do not result from code duplication, and pattern matching necessitates the expertise of human professionals to define vulnerability patterns. 

To address the limitation, neural networks have recently been employed for vulnerability detection~\citep{Yang2015,Shin2015,White2016,Wang2016,Li2017,Guo2017,Li2018VulDeePeckerAD,zhou2019devign,lin2020software}. Neural networks have gained widespread recognition in fields such as image processing and speech recognition, owing to their ability to deliver highly accurate predictions with minimal dependence on human experts for feature extraction. Given the diverse causes of software vulnerabilities, neural networks can be a valuable asset in their detection. Unlike pattern-based methods, neural networks automatically extract features and thus mitigate the impact of human bias in feature extraction.

This paper presents our work on using neural networks to create predictive models for automatically detecting vulnerabilities. It consists of four major steps: 1) extracting code relevant to vulnerabilities, 2) converting the extracted code into numeric vectors, 3) training and optimizing neural networks using the numeric vectors, and 4) detecting vulnerabilities using the models generated by the neural networks. 

First, we use program slicing to extract syntax and semantic information of four different types of program constructs relevant to vulnerabilities from the source code of target programs. The program constructs include library or API function call, array usage, pointer usage, and arithmetic expression. Each program slice contains the vulnerability-related program construct, and the program statements on which the program construct are control dependent or data dependent.

Second, the extracted program slices are then converted into numeric vectors using the Word2Vector model. Each slice is split into tokens, and the tokens are used as the input to the Word2Vector model, which learns word embedding and outputs the word embedding numeric vectors for the tokens.

Third, the numeric vectors representing the tokens are pre-processed and fed into neural networks. The purpose of the pre-processing is to improve the accuracy of the models generated by the neural networks in vulnerability detection. The pre-processing consists of dataset balancing and integration. 

We perform dataset balancing because our dataset has substantially more non-vulnerable program slices than vulnerable program slices, reflecting the fact that programs have much more non-vulnerable code than vulnerable code. To balance the dataset, we downsize the numeric vectors for non-vulnerable program slices. 

While prior work~\citep{li2018sysevr} trains individual models on each type of vulnerability-related program construct, our work trains on the data integrated from all types of program constructs. Our results demonstrate that the model built from the integrated dataset outperforms the individual models created from separate datasets.


To create a robust predictive model, we fine-tune the neural networks by experimenting with various hyperparameters, including optimizers and gating mechanisms, during model development. Our experiments show that the ADAM optimizer and bidirectional RNNs achieve the best results.

Lastly, we use the trained models to identify vulnerabilities in our dataset. Our BGRU model outperforms the BLSTM model. It achieves an accuracy rate of 94.6\% on the training set and 92.4\% on the test set.

The major contributions of this paper is as follows:
\begin{itemize}
    \item We show that the accuracy of the model built on the combined dataset surpasses the models built on individual dataset.
    \item By balancing the ratio of vulnerable data points (class 1) and non-vulnerable data points (class 0), the model performs well with a high balanced accuracy rate of 93\% which is comparable to that of a training set. The high sensitivity and specificity imply the model has a good ability in explaining both vulnerability and non-vulnerability classes. 
    \item We compare different types of neural networks and show that BGRU performs the best. The model built with BGRU achieves an accuracy rate of 94.89\% by utilizing 10X more data points. 
    \item We implement a chain of tools for generating the model from program slices and open source the tools at \url{https://gitlab.com/vulnerability_analysis/vulnerability_detection/}.
\end{itemize}

The paper is structured into six sections. Section~\ref{sec:dataset} presents information on the dataset. Section~\ref{sec:optimization} describes the details on fine-tuning the models. Section~\ref{sec:evaluation} shows evaluation results. Section~\ref{sec:related} discusses related work. Finally, we conclude in Section~\ref{sec:conclusion}. This paper expands upon the ideas presented in~\cite{AumpansubH21}.

\section{Dataset}\label{sec:dataset}
Our work uses the dataset of C/C++ programs collected by~\cite{li2018sysevr}. The dataset includes 1,592 programs from the National Vulnerability Database (NVD) and 14,000 programs from the Software Assurance Reference Dataset (SARD). These programs were pre-processed and transformed to 420,627 program slices called semantic vulnerability candidates (SeVC) which contain 56,395 vulnerable slices (13.5 \% of program slices) and 364,232 non-vulnerable slices (86.5 \% of program slices). The program slices are then transformed into numeric vectors that will be used as inputs to neural networks.

The program slices were created by extracting statements relevant to four types of vulnerability-relevant program constructs: 
\begin{itemize}
\item \scbf{Library or API Function Call (API)}
This type of program slices is associated with library or API functions calls for 811 C/C++ library/API function calls. This type represents 15.3\% of total slices, comprising 13,603 vulnerable slices and 50,800 non-vulnerable slices.

\item \scbf{Array Usage (AU)}
This type of program slices is related to the use of arrays such as array element access, accounting for 10\% of total slices which contain 10,926 vulnerable slices and 31,303 non-vulnerable slices.
      
\item \scbf{Pointer Usage (PU)}
This type of program slices is related to the use of pointer arithmetic and dereferences. This type represents 69.4\% of total slices which include 28,391 vulnerable slices and 263,450 non-vulnerable slices.

\item \scbf{Arithmetic Expression (AE)}
This type of program slices is associated with arithmetic expressions such as integer additions and subtractions, which represents 5.3\% of total slices, comprising 3,475 vulnerable slices  and 18,679 non-vulnerable slices.
\end{itemize}

\subsection{Generating Program Slices}
The program slices are generated in two phases. First, syntax-based vulnerability candidates (SyVCs) are extracted from programs, based on the abstract syntax trees (ASTs) of the programs. Each SyVC encapsulates the syntax characteristics of a vulnerability-related program construct. Second, semantics-based vulnerability candidates (SeVCs) are generated from SyVCs by generating a program dependency graph (PDG) for each function of the programs and extending each SyVC with data dependency and control dependency information from PDGs. Each SeVC is a program slice that contains semantic and syntax information related with a vulnerability-related program construct. We define the type of a program slice as the type of program construct on which the program slice is generated.

The process of generating program slices is illustrated in Figure~\ref{fig:preprocessing}. The Joern package in Python was used to parse the source code and generate PDG. More details on program slice generation can be found in~\cite{li2018sysevr}.

\myfig{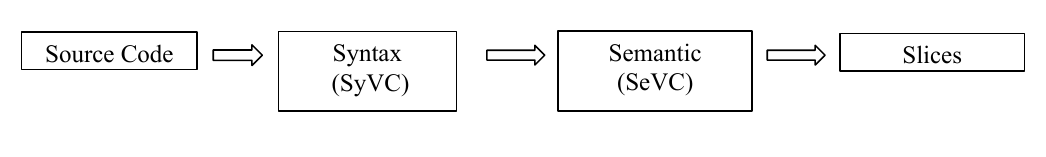}{Generating Program Slices fro Source Code. }{fig:preprocessing}{1.0}{0pt}

\subsection{Transforming Program Slices into Vectors}
To use the program slices with neural networks, they need to be transformed into numeric vectors. Each slice is first split into a list of tokens in which all comments and white spaces were removed. It is also mapped to the list of relevant functions.


The list of tokens for each slice are stored in a pickle file and labeled with a unique ID. Each pickle file contains five elements: a list of tokens, a target label (0/1), a list of functions, vulnerability type, and the ID of the slice. A target label of 0 indicates that the slice is non-vulnerable, while a target label of 1 indicates that the slice is vulnerable.

The list of tokens from each pickle file is converted intto vectors using the Word2Vector model, which converts tokens to vectors based on cosine similarity distance, measuring the angle between vectors. A higher similarity score indicates a higher similarity and a closer distance between tokens~\citep{mikolov2013efficient}. The cosine similarity is computed as follows:

\myfig{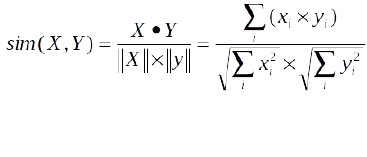}{Cosine similarity.}{}{0.5}{0pt}

For each program slice, the output of the Word2Vector model is a $30 \times n$ array, where 30 is the dimension of the columns and $n$ is the dimension of the rows. Each row is the word embedding for one token and thus $n$ is the number of tokens in the program slice.

The visualization of tokens in the Word2Vector model is shown in Figure~\ref{fig:visualization}. As we can see, different program slice types have substantially different distributions of cosine similarities. This indicate that different program slice types convey different characteristics of vulnerabilities.

\myfig{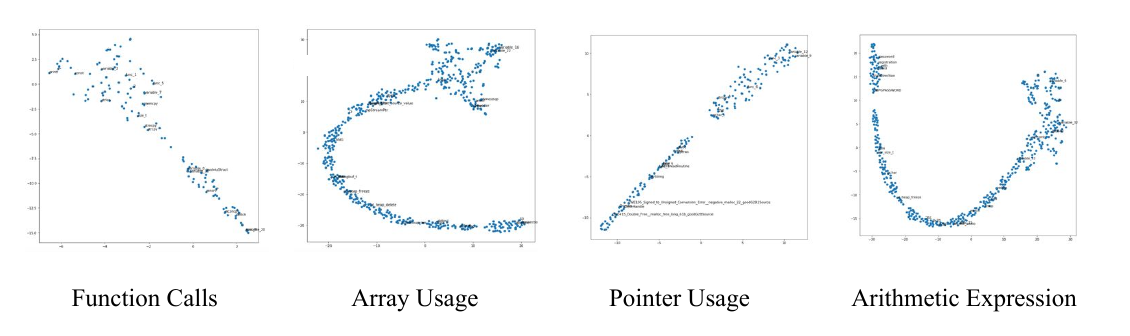}{Visualized tokens in W2V model for each program slice type.}{fig:visualization}{1.0}{0pt}

\section{Model Optimization}\label{sec:optimization}
In this section, we describe the steps that we took to find optimal pre-processing techniques and neural network models. We use a subset of the dataset for the majority of our experiments. The subset includes randomly chosen 30,000 vector arrays from the total 420,627 vector arrays, each of which corresponds to a program slice. The subset is split into a training set of 24,000 vector arrays and a testing set of 6,000 vector arrays. First, we compare the results on individual program slice types and the results on combined program slice types. Second, we show the results using an imbalanced dataset and the results using a balanced dataset. Third, we experiment with various optimizers including SGD, ADAMAX, and ADAM. Last, we discuss and compare the results using different RNNs.

\subsection{Combining Program Slice Types}
From the visualization of Word2Vector models for each program slice type, as shown in Figure~\ref{fig:visualization}, we note that different program slice types capture different characteristics of vulnerabilities, so we explore the use of a dataset combined from all different program slice types. 

We perform a preliminary study using 1,000 randomly chosen program slices from each individual program slice types, collectively called individual datasets, and 1,000 randomly chosen program slices from all different program slice types, called combined dataset. We compare the accuracy, sensitivity, and specificity for the models built using individual datasets and the model built using the combined dataset. Table~\ref{tbl:combination} shows our result.

\begin{table}[ht]
    \centering
    \caption{Comparison between individual datasets and combined dataset.}
    \label{tbl:combination}    
    \begin{tabular}{|c|r|r|r|}
        \hline
       Type & Accuracy & Sensitivity & Specificity \\
        \hline
         API & 53\% & 69\% & 46\%  \\
        \hline
         AU &  64\% & 79\% & 62\% \\
         \hline        
         PU & 38\% & 83\% & 31\% \\
         \hline         
         AE & 61\% & 61\% & 62\% \\
         \hline         
         COMBINED & 61\% & 91\% & 53\% \\
         \hline
    \end{tabular}
\end{table}

The model built using the combined dataset, i.e. combined model, outperforms the models built using individual datasets, i.e. individual models, in detecting the target class 1 (vulnerable) code, as the sensitivity of the model is 91\%, the highest among the all models. In detecting the target class 0 (non-vulnerable) code, the combined model performs considerably better than the API model and PU model, while it performs slightly worse than the AU model and AE model. As a result, we consider the combined dataset more appropriate for predicting vulnerabilities. 

\subsection{Balancing Dataset}
In general, the dataset used for training should have a balanced number of class 0 samples (non-vulnerable code) and class 1 samples (vulnerable code) to ensure that the model can produce unbiased predictions. However, ~\cite{li2018sysevr} used an imbalanced dataset in which class 1 samples only account for 15.6\% of the total program slices while class 0 samples account for 84.4\% of the total program slices. 

To illustrate the issue, we compute the confusion matrix for the model built with imbalanced dataset (75\% of class 0 and 25\% of class 1). As presented in Table~\ref{tbl:imblanced}, the model has considerably higher accuracy in predicting class 0 samples, as its specificity and negative prediction are remarkably higher than its sensitivity and precision, respectively. Its accuracy rate is biased towards class 0. 

\begin{table}[ht]
    \centering
    \caption{Confusion matrix for imbalanced dataset.}
    \label{tbl:imblanced}
    \begin{tabular}{|c|l|l|l|l|}
        \hline
        Predicted Class & & & & \\
        \hline
        \hline
                 & Positive & Negative & Rate & \\
        \hline
        Positive & 8.0 & 48.0 & 0.14285 & Sensitivity \\  
        \hline
        Negative & 14.0 & 130.0 & 0.90277 & Specificity \\
        \hline
                 & 0.36363 & 0.73033 & 0.60999 & Accuracy \\
        \hline
                & Precision & Negprediction & & \\
        \hline
    \end{tabular}
\end{table}

In order to address the issue, we re-sample the training set using a down-sampling method, which randomly removes samples from the majority class (label 0) of the training set to make the number of class 0 samples the same as the number of class 1 samples. The new training set has a balanced samples, containing 50\% vulnerable samples and 50\% non-vulnerable samples. Figure~\ref{fig:downsampling} shows the process of down-sampling. 

Because neural networks require all vector input to have the same dimension, we also adjust our vector arrays to have the same number of rows, i.e. same vector lengths. We compute the average number of rows of the vector arrays, and use it as the threshold to adjust the vector lengths. If a vector array has a vector length less than the average length, we append the vector array with zero vectors. If a vector array has a vector length larger than than the average length, we truncate the vector array. 

\myfig{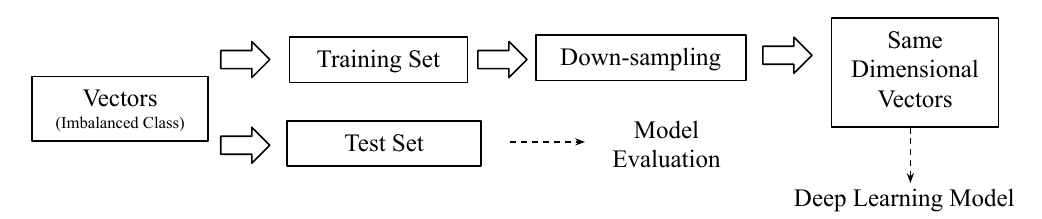}{Down-sampling and vector adjustment.}{fig:downsampling}{1.0}{0pt}

As shown in Table~\ref{tbl:blanced}, the model built using the balanced dataset has approximately the same accuracy in predicting class 0 samples and class 1 samples. This shows that balancing the dataset is critical for the model to have balanced prediction power for both classes.

\begin{table}[ht]
    \centering
    \caption{Confusion matrix for balanced dataset.}
    \label{tbl:blanced}
    \begin{tabular}{|c|l|l|l|l|}
        \hline
        Predicted Class & & & & \\
        \hline
        \hline
                 & Positive & Negative & Rate & \\
        \hline
        Positive & 1186.0 & 167.0 & 0.87916 & Sensitivity \\  
        \hline
        Negative & 672.0 & 4018.0 & 0.86408 & Specificity \\
        \hline
                 & 0.65236 & 0.96108 & 0.86747 & Accuracy \\
        \hline
                & Precision & Negprediction & & \\
        \hline
    \end{tabular}
\end{table}

\subsection{Selecting Optimizers}
Different optimizers can be applied to optimize neural networks. They are algorithms for finding the optimal parameters for a model during the training process by adjusting the weights and biases in the model iteratively until they converge on a minimum loss value. Some of the most popular optimizers include SGD, Momentum, ADAMGRAD, RMS Prop, ADAM and ADAMAX. In order to find the best optimizer for our neural networks, we explore three different optimizers: SGD, ADAM, and ADAMAX. 

SGD computes the gradient of the loss function based on a randomly chosen subset of the training data instead of the entire training data. Comparing to the standard gradient descent, SGD can converge faster and use less memory storage.

ADAM computes individual learning rates for different parameters. It keeps track of a changing average of the gradient's first and second moments, which are respectively the mean and variance of the gradients. ADAM is appropriate for large dataset and/or parameters, with  non-stationary objectives, and for problems with very noisy and/or sparse gradients~\cite{Kingma:2014vow}. 

ADAMAX is a variant of ADAM. Similar to ADAM, ADAMAX also keeps track of a changing average of the gradient's mean and variance of the gradients. Different from ADAM, ADAMAX uses the L-infinity norm of the gradients instead of the second moment of the gradients. ADAMAX is appropriate for the scenarios in which the gradients are sparse or have a high variance.

\begin{table}[]
    \centering
    \caption{Accuracy rate with different optimizers.}
    \label{tbl:optimizers}
    \begin{tabular}{|c|r|r|r|}
        \hline
         Type & ADAMAX & SGD & ADAM \\
                 \hline
         API & 86.7\% & 63.1\% & 89.5\% \\
                 \hline
         AU & 86.0\% & 58.6\% & 89.2\% \\
                 \hline
         PU & 82.4\% & 62.3\% & 90.9\% \\
                 \hline
         AE & 83.1\% & 67.1\% & 90.5\% \\
        \hline
    \end{tabular}
\end{table}

The accuracy of the models using ADAMAX, SGD, and ADAM optimizer is presented in Table~\ref{tbl:optimizers}. We can see that ADAM performs the best among the three optimizers for all program slice types. ADAM achieves an average accuracy rate of 90.0\%. This is approximately 5\% higher than the accuracy rate of ADAMAX, which is used in prior work~\citep{li2018sysevr}. As a result, we choose to use ADAM for our neural networks.

\subsection{Comparing RNNs}
In this section, we discuss and compare the performance of different neural networks. First, we discuss GRU and LSTM. Second, we compare LSTM with BLSTM. Last, we analyze BGRU and BLSTM.

\scbf{GRU vs. LSTM} Comparing to LSTM, GRU has no explicit memory unit, no forget gate and update gate. GRU also has fewer number of hyperparameters. With a simpler architecture, GRU trains faster than LSTM. However, GRU may have lower accuracy rate than LSTM, because LSTM comprises both update gate and forget gate and remembers longer sequences than GRU, although LSTM is comparable to GRU on sequence modeling.

\scbf{BLSTM vs. LSTM} A bidirectional recurrent neural network (RNN) has two layers side-by-side. It provides the original input sequence to the first layer and a reversed copy of the input sequence to the second layer. Bidirectional RNNs are found to be more effective than regular RNNs, because it can overcome the limitations of a regular RNN~\citep{Schuster1997}. A regular RNN preserves only information of the past, while a bidirectional RNN has access to the past information as well as the future information. Therefore the output of a bidirectional RNN is generated from both the past context and future context, and that leads to a better prediction and classifying capability. Our experiment in training LSTM and BLSTM models on a subset of our dataset also indicates that BLSTM outperforms LSTM using the same hyperparameters, as shown in Figure~\ref{fig:BLSTM}. 

\myfig{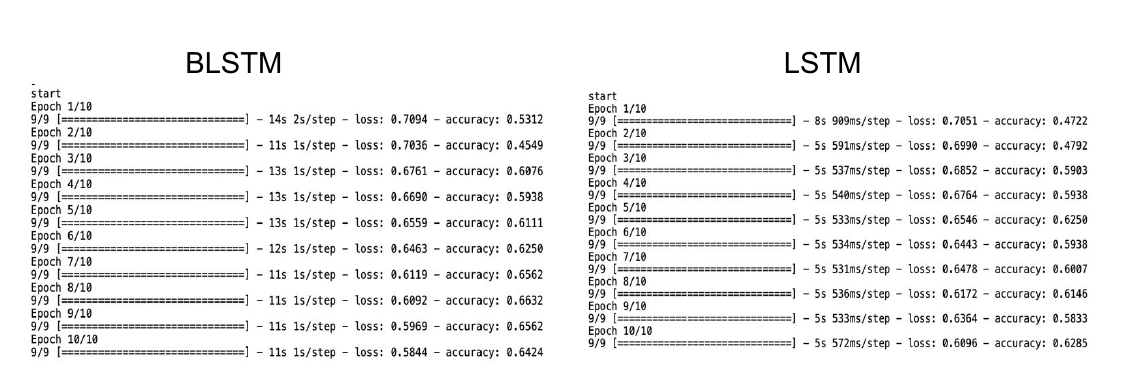}{Model fitting of BLSTM and LSTM.}{fig:BLSTM}{1.0}{0pt}

The result of this experiment shows that the BLSTM model has a lower loss rate of 0.58, as compared to the LSTM model's loss rate of 0.60. The BLSTM model also has a higher accuracy rate of 64.2\% than the LSTM model's accuracy rate of 62.8\%. Note that in this experiment both models were fit with the same input parameters on a small dataset, which includes 1,000 program slices, so the accuracy rates are not high. 

\begin{table}[ht]
    \centering
    \caption{Confusion matrix for BGRU with 5,000 samples.}
    \label{tbl:bgru_5000}
    \begin{tabular}{|c|l|l|l|l|}
        \hline
        Predicted Class & & & & \\
        \hline
        \hline
                 & Positive & Negative & Rate & \\
        \hline
        Positive & 185.0 & 41.0 & 0.81858 & Sensitivity \\  
        \hline
        Negative & 295.0 & 478.0 & 0.61837 & Specificity \\
        \hline
                 & 0.38542 & 0.92100 & 0.66366 & Accuracy \\
        \hline
                & Precision & Negprediction & & \\
        \hline
    \end{tabular}
\end{table}

\begin{table}[ht]
    \centering
    \caption{Confusion matrix for BLSTM with 5,000 samples.}
    \label{tbl:blstm_5000}
    \begin{tabular}{|c|l|l|l|l|}
        \hline
        Predicted Class & & & & \\
        \hline
        \hline
                 & Positive & Negative & Rate & \\
        \hline
        Positive & 185.0 & 41.0 & 0.81858 & Sensitivity \\  
        \hline
        Negative & 321.0 & 452.0 & 0.58473 & Specificity \\
        \hline
                 & 0.36561 & 0.91684 & 0.63764 & Accuracy \\
        \hline
                & Precision & Negprediction & & \\
        \hline
    \end{tabular}
\end{table}

\scbf{BGRU vs. BLSTM} Table~\ref{tbl:bgru_5000} and Table~\ref{tbl:blstm_5000} show the confusions matrices for BGRU and BLSTM respectively. The models were trained on 4,000 samples and tested on 1,000 samples. The decision threshold is set to 0.5 for validation. The BGRU model outperforms the BLSTM in most metrics except for the sensitivity. It has a higher accuracy, precision, and specificity, which indicates its stronger capability to predict both vulnerable code and non-vulnerable code. 

\begin{table}[ht]
    \centering
    \caption{Confusion matrix for BGRU with 30,000 samples.}
    \label{tbl:bgru_30000}
    \begin{tabular}{|c|l|l|l|l|}
        \hline
        Predicted Class & & & & \\
        \hline
        \hline
                 & Positive & Negative & Rate & \\
        \hline
        Positive & 731.0 & 77.0 & 0.90470 & Sensitivity \\  
        \hline
        Negative & 1022.0 & 4170.0 & 0.80316 & Specificity \\
        \hline
                 & 0.41699 & 0.98187 & 0.81683 & Accuracy \\
        \hline
                & Precision & Negprediction & & \\
        \hline
    \end{tabular}
\end{table}

\begin{table}[ht]
    \centering
    \caption{Confusion matrix for BLSTM with 30,000 samples.}
    \label{tbl:blstm_30000}
    \begin{tabular}{|c|l|l|l|l|}
        \hline
        Predicted Class & & & & \\
        \hline
        \hline
                 & Positive & Negative & Rate & \\
        \hline
        Positive & 663.0 & 145.0 & 0.82054 & Sensitivity \\  
        \hline
        Negative & 1095.0 & 4097.0 & 0.78910 & Specificity \\
        \hline
                 & 0.37713 & 0.96582 & 0.79333 & Accuracy \\
        \hline
                & Precision & Negprediction & & \\
        \hline
    \end{tabular}
\end{table}

BGRU outperforms BLSTM in all the metrics on a larger dataset of 30,000 samples. As shown in Table~\ref{tbl:bgru_30000} and Table~\ref{tbl:blstm_30000}, the BGRU model has a sensitivity of 90\%, which is 8\% higher than that of the BLSTM model. This indicates that the BGRU model can predict the vulnerable code better than the BLSTM model. Table~\ref{tbl:bgru_100000} and Table~\ref{tbl:blstm_100000} show that the BGRU model also performs better than the BLSTM model on a dataset of 100,000 samples. Comparing to the BLSTM model, the BGRU model has 3\% higher accuracy and specificity, and approximately the same sensitivity.

\begin{table}[ht]
    \centering
    \caption{Confusion matrix for BGRU with 100,000 samples.}
    \label{tbl:bgru_100000}
    \begin{tabular}{|c|l|l|l|l|}
        \hline
        Predicted Class & & & & \\
        \hline
        \hline
                 & Positive & Negative & Rate & \\
        \hline
        Positive & 2383.0 & 287.0 & 0.89251 & Sensitivity \\  
        \hline
        Negative & 1506.0 & 15824.0 & 0.91310 & Specificity \\
        \hline
                 & 0.61275 & 0.98219 & 0.91035 & Accuracy \\
        \hline
                & Precision & Negprediction & & \\
        \hline
    \end{tabular}
\end{table}

\begin{table}[ht]
    \centering
    \caption{Confusion matrix for BLSTM with 100,000 samples.}
    \label{tbl:blstm_100000}
    \begin{tabular}{|c|l|l|l|l|}
        \hline
        Predicted Class & & & & \\
        \hline
        \hline
                 & Positive & Negative & Rate & \\
        \hline
        Positive & 2408.0 & 262.0 & 0.90187 & Sensitivity \\  
        \hline
        Negative & 2193.0 & 15137.0 & 0.87346 & Specificity \\
        \hline
                 & 0.52336 & 0.98300 & 0.87725 & Accuracy \\
        \hline
                & Precision & Negprediction & & \\
        \hline
    \end{tabular}
\end{table}


\section{Evaluation}\label{sec:evaluation}
In this section, we evaluate the accuracy of our neural networks in predicting vulnerable code and non-vulnerable code. We first show the results on individual program slice types, then show the results on the combined program slice types. We build the models using BGRU for all the evaluations.

\subsection{Individual Program Slice Types}
The dataset contains program slices created for four types of vulnerability-related program constructs: library or API functions (API), array usage (AU), pointer usage (PU), and arithmetic expressions (AE). We build individual models for each program slice type. We use a dataset of 6,000 program slices for this evaluation. Our focus is to find a threshold on the prediction results that will have the best accuracy.

Table~\ref{tbl:threshold} presents the threshold for the BGRU model to achieve the highest accuracy rate and F1 score for each type of program slices. As we can, the threshold to get the highest accuracy rate ranges from 0.4 to 0.65, with a mean of 0.5, and the threshold to get the highest F1 score ranges from 0.55 to 0.7, with a mean of 0.625.

\begin{table}[ht]
    \centering
    \caption{Prediction thresholds for different program slice types.}
    \label{tbl:threshold}
    \begin{tabular}{|l|l|l|}
        \hline
        Type & Threshold for Accuracy & Threshold for F1\\
        \hline
        API & 0.55 & 0.65\\
        \hline
        AU & 0.4 & 0.55\\
        \hline
        PU & 0.4 & 0.6\\
        \hline
        AE & 0.65 & 0.7\\
        \hline
    \end{tabular}
\end{table}

\scbf{API} The effects of different thresholds for API slices are shown in Table~\ref{tbl:API_threshold}. A prediction threshold of 0.55 achieves the highest accuracy of 0.872 while a predication threshold of 0.65 achieves the highest F1 score of 0.759.

\begin{table}[ht]
    \centering
    \caption{Prediction threshold for API slices.}
    \label{tbl:API_threshold}
    \begin{tabular}{|l|l|l|l|l|l|}
        \hline
        Threshold & Recall & Precision & Specificity & F1 & Accuracy\\
        \hline
        0.4 & 0.906597 & 0.617677 & 0.837204 & 0.734755 & 0.871901 \\
        \hline
        0.45 & 0.889548 & 0.630252 & 0.848602 & 0.737781 & 0.869075 \\
        \hline
        0.5 & 0.87917 & 0.652365 & 0.864086 & 0.748974 & 0.871628 \\
        \hline
        0.55 & 0.870274 & 0.668184 & 0.874624 & 0.755956 & 0.872449 \\
        \hline
        0.6 & 0.858414 & 0.678781 & 0.882151 & 0.758101 & 0.870282 \\
        \hline
        0.65 & 0.841464 & 0.690809 & 0.890753 & 0.75869 & 0.866058 \\
        \hline
        0.7 & 0.811712 & 0.702824 & 0.90043 & 0.753354 & 0.856071 \\
        \hline
    \end{tabular}
\end{table}

\scbf{Array Usage} The effects of different thresholds for AU slices are shown in Table~\ref{tbl:AU_threshold}, a prediction threshold of 0.4 achieves the highest accuracy of 0.878 while a predication threshold of 0.55 achieves the highest F1 score of 0.812.

\begin{table}[ht]
    \centering
    \caption{Prediction threshold for Array Usage slices.}
    \label{tbl:AU_threshold}
    \begin{tabular}{|l|l|l|l|l|l|}
        \hline
        Threshold & Recall & Precision & Specificity & F1 & Accuracy\\
        \hline
        0.4 & 0.946099 & 0.704511 & 0.809183 & 0.807625 & 0.877641 \\
        \hline
        0.45 & 0.931725 & 0.713724 & 0.820291 & 0.808283 & 0.876008 \\
        \hline
        0.5 & 0.918891 & 0.724696 & 0.83214 & 0.810321 & 0.856031 \\
        \hline
        0.55 & 0.904517 & 0.737238 & 0.844977 & 0.812356 & 0.874747 \\
        \hline
        0.6 & 0.86961 & 0.745599 & 0.857319 & 0.802844 & 0.86131 \\
        \hline
        0.65 & 0.826489 & 0.770704 & 0.881758 & 0.797622 & 0.854123 \\
        \hline
        0.7 & 0.766427 & 0.800966 & 0.908418 & 0.783316 & 0.837422 \\
        \hline
    \end{tabular}
\end{table}

\scbf{Pointer Usage} The effects of different thresholds for PU slices are shown in Table~\ref{tbl:PU_threshold}, a prediction threshold of 0.4 achieves the highest accuracy of 0.837 while a predication threshold of 0.6 achieves the highest F1 score of 0.693.

\begin{table}[ht]
    \centering
    \caption{Prediction threshold for Pointer Usage slices.}
    \label{tbl:PU_threshold}
    \begin{tabular}{|l|l|l|l|l|l|}
        \hline
        Threshold & Recall & Precision & Specificity & F1 & Accuracy\\
        \hline
        0.4 & 0.885281 & 0.556715 & 0.788207 & 0.683565 & 0.836744 \\
        \hline
        0.45 & 0.872294 & 0.568139 & 0.80078 & 0.688105 & 0.836537 \\
        \hline
        0.5 & 0.844156 & 0.582669 & 0.818339 & 0.689452 & 0.831248 \\
        \hline
        0.55 & 0.822511 & 0.599054 & 0.834598 & 0.69322 & 0.828554 \\
        \hline
        0.6 & 0.798701 & 0.612618 & 0.848255 & 0.693392 & 0.823478 \\
        \hline
        0.65 & 0.771284 & 0.624051 & 0.860395 & 0.6899 & 0.815839 \\
        \hline
        0.7 & 0.731602 & 0.641366 & 0.877086 & 0.683519 & 0.804344 \\
        \hline
    \end{tabular}
\end{table}

\scbf{Arithmetic Expression} The effects of different thresholds for AE slices are shown in Table~\ref{tbl:AE_threshold}, a prediction threshold of 0.65 achieves the highest accuracy of 0.878 while a predication threshold of 0.7 achieves the highest F1 score of 0.677.

\begin{table}[ht]
    \centering
    \caption{Prediction threshold for Arithmetic Expression slices.}
    \label{tbl:AE_threshold}
    \begin{tabular}{|l|l|l|l|l|l|}
        \hline
        Threshold & Recall & Precision & Specificity & F1 & Accuracy\\
        \hline
        0.4 & 0.936324 & 0.444979 & 0.784167 & 0.603263 & 0.860246 \\
        \hline
        0.45 & 0.930535 & 0.46259 & 0.800214 & 0.617972 & 0.865375 \\
        \hline
        0.5 & 0.927641 & 0.479073 & 0.813587 & 0.631838 & 0.870614 \\
        \hline
        0.55 & 0.920405 & 0.496487 & 0.827494 & 0.64503 & 0.87395 \\
        \hline
        0.6 & 0.914616 & 0.514239 & 0.840332 & 0.658333 & 0.877474 \\
        \hline
        0.65 & 0.901592 & 0.53339 & 0.854239 & 0.670253 & 0.877915 \\
        \hline
        0.7 & 0.885673 & 0.548387 & 0.865205 & 0.677366 & 0.875439 \\
        \hline
    \end{tabular}
\end{table}

\subsection{Combined Program Slice Types}
We combine the total 420,067 programs slices into one dataset, comprising 64,403, 42,229, 291,281, and 22,154 from API, AU, PU, and AE types, respectively. The combined dataset is split into a training set and a test set with the 80:20 ratio. The training set is then down-sampled to ensure that the target classes (vulnerable and non-vulnerable) in it are balanced. 

Our BGRU model is built with the ADAM optimizer. The hyperparameters of the model include 256 neuron units with 2 hidden layers. The Tanh function is applied to produce the outputs of 2 hidden layers and the Sigmoid function is applied to compute activation outputs in the last layer. The learning rate is 0.1 with a batch size of 32. The binary cross-entropy loss function is used as it can speed up the convergence. 

We illustrates the learning process in Figure~\ref{fig:modelfitting}. The learning process is faster in the beginning, as the loss rate significantly decreases in epoch 1 to 3. The accuracy rate increases as the training process goes from epoch 1 to 10. The model has the highest accuracy rate of 94.89\% in epoch 9 and starts to decrease in epoch 10 as the error rate is no longer reduced. The output of the model ranges between 0 and 1, as the Sigmoid function is applied to the output layer. 

\myfig{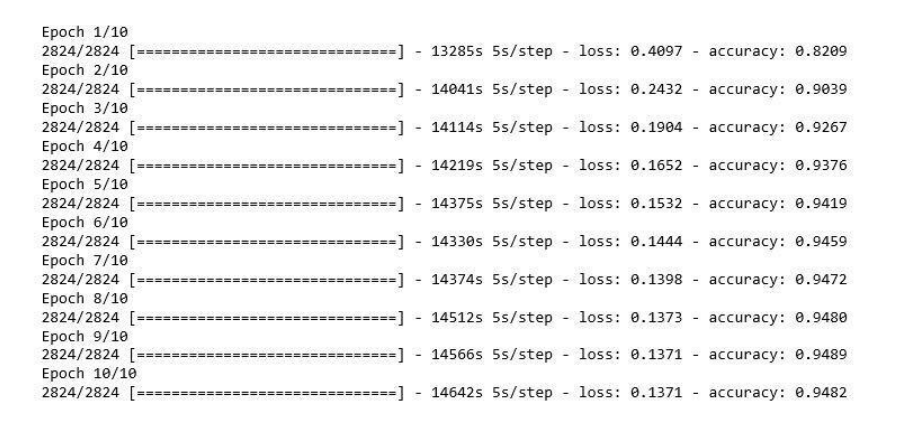}{Model fitting with training set.}{fig:modelfitting}{1.0}{0pt}

\begin{table}[h]
    \centering
    \caption{Confusion matrix for test set.}
    \label{tbl:testset}
    \begin{tabular}{|c|l|l|l|l|}
        \hline
        Predicted Class & & & & \\
        \hline
        \hline
                 & Positive & Negative & Rate & \\
        \hline
        Positive & 10768.0 & 439.0 & 0.96082 & Sensitivity \\  
        \hline
        Negative & 5898.0 & 67019.0 & 0.91911 & Specificity \\
        \hline
                 & 0.64610 & 0.99349 & 0.92467 & Accuracy \\
        \hline
                & Precision & Negprediction & & \\
        \hline
    \end{tabular}
\end{table}

Table~\ref{tbl:testset} shows the confusion matrix for the test set. We can see that the model performs well in predicting both target classes (vulnerable code and non-vulnerable code), as both the sensitivity and specificity are over 90\%. 

As presented in Figure~\ref{fig:threshold}, the F1 score increases and the balanced accuracy decreases while the threshold increases. The peak point of the balanced accuracy is achieved when the threshold is 0.5. The peak point of the F1 score is achieved when the threshold is 0.8. 

\myfig{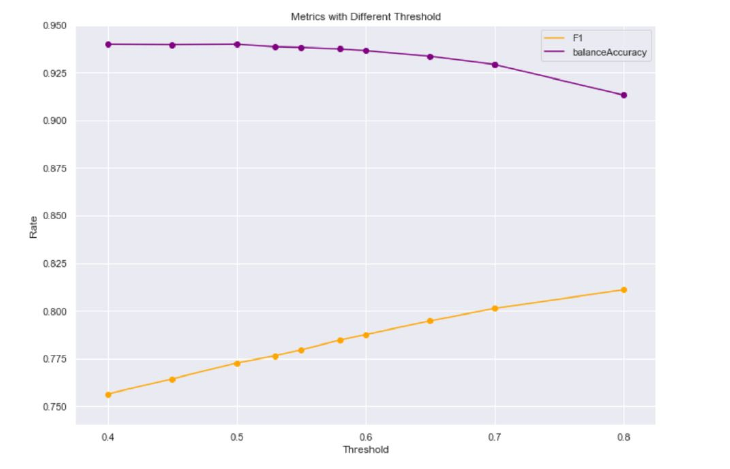}{F1 v.s. Accuracy Rate for Different Thresholds.}{fig:threshold}{1.0}{0pt}

Overall, the model fitted with the combined dataset performs well with a high accuracy rate of 92.5\%. Its high sensitivity and specificity indicates that it has a good capability in predicting both vulnerable code and non-vulnerable code, although the model performs better in predicting non-vulnerable code than vulnerable code, as it has a negative prediction rate of 99.3\%. 

\section{Related Work}\label{sec:related}


Many approaches have been proposed to detect and address vulnerabilities~\citep{valeur2005learning,Neuhaus2007,dessiatnikoff2011clustering,Shin2011,Yamaguchi2012,zheng2013path,Ocasta,VDiscover,VulPecker,Wu2017,SAIC,david2018firmup,Li2018VulDeePeckerAD,li2018sysevr,Chernis2018,wang2010taintscope,RVM,Li2019,lin2020software,Zagane2020,li2021sysevr,HuangY21,li2021vulnerability,Huang2021EuroSec,eshghie2021dynamic,hin2022linevd,CSR2022,fu2022linevul,Aumpansub2022,cao2022mvd}. They can be broadly categorized as rule-based approaches and learning-based approaches.

Rule-based approaches detect the existence of vulnerabilities using predefined rules, which typically characterize vulnerable and non-vulnerable program code structures~\citep{david2018firmup} or behaviors~\citep{wang2010taintscope}. Rule-based approaches follow the predefined rules to analyze program code or program behaviors. These analyses can be performed dynamically~\citep{HuangY21}, which execute target programs, or statically~\citep{zheng2013path}, which examine target programs without executing them. Rule-based approaches identify a vulnerability when a predefined rule finds a match of vulnerable code structures or behaviors. A major disadvantage of rule-based approaches is that the predefined rules require considerable manual effort and time to generate.

As learning-based approaches have thrived in a myriad of areas, particularly in software security and reliability~\citep{snitch,yuan2011context,Ocasta,Grieco2016,Wang2016,Long2016,SAIC,Petoumenos2018,Li2019,Tien2020}, they have also been leveraged in vulnerability detection. Learning-based approaches extract characteristics of program code or behaviors automatically and identify vulnerabilities based on these characteristics. Conventional machine learning approaches learn characteristics of vulnerabilities using various human-defined features such as source code text features in the source code ~\citep{Chernis2018}, complexity, code churn, and developer activity metrics~\citep{Shin2011}, abstract syntax trees~\citep{Yamaguchi2012}, function imports and function calls~\citep{Neuhaus2007}. Similar to rule-based approaches, a main drawback of conventional machine learning approaches is that they require considerable human effort to define these features.

Recent approaches use deep learning on program code to detect vulnerabilities so that no human experts is needed to define features~\citep{Li2018VulDeePeckerAD,li2018sysevr,Li2019,Zagane2020,li2021sysevr,hin2022linevd,li2021vulnerability,fu2022linevul,Aumpansub2022,cao2022mvd}. They typically use neural networks to automatically build classification models from a large number of program samples. Deep learning approaches have been shown to have better accuracy than conventional machine learning approaches~\citep{Wu2017}. However, most of them either rely on one type of training data or use imbalanced training data. Our work differs from them by using a balanced dataset that combines different types of training data. 


\section{Conclusion}\label{sec:conclusion}
We present our work on detecting software vulnerabilities using neural networks. In this work, we train neural networks with program slices extracted from the source code of 15,592 C/C++ programs. The program slices encapsulate  characteristics of different types of vulnerability-related program constructs. We compare different types of training data and different types of neural networks. Our results show that the model based on the combined slices of different program construct types outperforms the models based on the slices of individual program construct types. Using a balanced number of vulnerable program slices and non-vulnerable program slices ensures that the model has a balanced accuracy in predicting both vulnerable code and non-vulnerable code. We find that BGRU performs the best among other neural networks. It achieves an accuracy of 94.89\%, with a sensitivity of 96.08\% and a specificity of 91.91\%.
\bibliographystyle{agsm}
\bibliography{bibfile}


\end{document}